\documentstyle[12pt]{article}

\begin{document}
\begin{titlepage}
\vskip 1cm
\begin{flushright}
Preprint SPBU--IP--98--9
\end{flushright}
\begin{center}
{\Large\bf Effective Theories \\ with Maximal Analyticity}
\vskip 2cm
{\Large Alexander V. Vereshagin,\ \ \ \
        Vladimir V. Vereshagin
\footnote{{\it e-mail}:
{\tt vvv@av2467.spb.edu}}}
\vskip 1cm
{\it Theoretical Physics Department\\
St.Petersburg State University\\
Ulyanovskaya st 1\\
198904 St.Petersburg\\
Russia}\\[12pt]
\end{center}
\vfill

\begin{abstract}
\normalsize
In this paper (second in the series) we study the properties
of tree-level binary amplitudes of the infinite-component
effective field theory of strong interaction obeying the
requirements of quark-hadron duality and maximal analyticity.
In contrast to the previous paper, here we derive the results
following from less restrictive --- Regge-like --- boundedness
conditions. We develop the technique of Cauchy's forms in two
variables and show the string-like structure of a theory. Next,
we derive the full set of bootstrap constraints for the
resonance parameters in $(\pi,K)$ system. Numerical test shows:
(1) those constraints are consistent with data on well
established vector resonances; (2) two light broad resonances
-- $\sigma$ and $\kappa$ -- are needed to saturate sum rules
following from Chiral symmetry and analyticity. As a by-product
we obtain expressions for the parameters of Chiral expansions
and give corresponding estimates.

\end{abstract}

\end{titlepage}

\section{Introduction}
\mbox{}

In a previous paper by one of us \cite{1} it was shown that the
requirements of meromorphy and polynomial boundedness applied
to the most general form of tree-level amplitude of a given
binary process give rise to a certain infinite system of
constraints for coupling constants and particle masses. Also it
was pointed out that those constrains realize the so-called
bootstrap conditions which -- in turn -- mirror the dual
properties of hadronic amplitudes.

Some of the bootstrap conditions for $(\pi, K)$ scattering
amplitude derived in ref \cite{1} can be checked numerically
because the modern experimental data provide the necessary
values for masses and coupling constants. Since the conditions
in question take a form of rapidly converging sum rules (SR),
one can select for numerical check those SR which can be
saturated (with sufficient accuracy) by the contributions of
few lightest resonances.

After the checking of several SR we have recognized the presence
of systematical discrepancies between their left and right hand
sides; those discrepancies could be hardly explained by
incompleteness of modern database. The point is that -- roughly
speaking -- our analysis has shown that the relative magnitude of
the two most significant contributions (those of $\rho-$ and
$K^{*}-$ mesons) following from SR derived in \cite{1} were
inconsistent with the well established values of the
corresponding masses and coupling constants.

This observation shows that the system of postulates accepted in
\cite{1} is inconsistent with the physical reality and, hence, it
must be reconsidered. This is done in a given paper. It is shown
that replacement of the decreasing asymptotics requirement for
the inelastic channel amplitude by the Regge conditions results
in a new system of bootstrap constrains which is quite reasonable
from the phenomenological point of view.

The paper is organized as follows. In Sec.~2 we explain the
essence of the actual physical problem which general solution
this paper is devoted to. Besides, we explain also the
constructive formulation of the maximal analyticity principle
which plays a key role in our approach. Sec.~3 is the central
one: here we give the general outline of the mathematical tool
specially constructed to work with meromorphic functions of
two (and more) variables with fiberwise given asymptotics. Since
in \cite{1} it is shown that our approach gives rise to certain
duality properties, in Sec.~4 we consider the widely known
example of the dual (string) amplitude constructed from a single
$B-$function. This analysis allows us to point out some
particular suggestions implicitly contained in conventional dual
hadron amplitudes. In Sec.~5 we apply the developed technique to
derive the set of bootstrap constraints for the parameters of
$(\pi ,K)$ amplitude, this set basing on weaker suggestions
(compared to those accepted in \cite{1}) about the asymptotics of
inelastic channel. In contrast with \cite{1} we show here the
explicit form of generating functions allowing us to write down
(in Sec.~6) several rapidly converging bootstrap conditions (sum
rules) which can be easily checked with the help of known data on
spectrum parameters. After the checking of those SR validity, in
Sec.~7 we derive explicit expressions for low energy parameters
and compute the corresponding numerical values. Besides, we show
that our SR require the existence of light scalar resonances with
isospins $I=0,1/2$ and estimate their parameters. At last, Sec.~8
is devoted to the discussion of the results obtained. Appendix
contains the necessary formulae and relations.

\section{Preliminary notes}
\mbox{}

First of all we would like to recall the essence of the problem
which stimulated us to begin a systematic study of the
properties of tree-level amplitudes in a framework of effective
field theory approach. This is the widely discussed problem of
low energy coefficients (LEC's) appearing in chiral expansions
\cite{3,4,5} (the excellent discussion can be found in \cite{6}).
Those coefficients cannot be fixed with a help of the symmetry
constraints since they are nothing but coupling constants
corresponding to various invariant interaction lagrangians. The
number of LEC's very rapidly increases with the expansion order.
This very circumstance creates a problem because it reduces to
zero the predictive power of Chiral perturbation theory (ChPT) in
higher orders. That is why it would be very interesting to find a
way allowing one to fix the LEC's or, at least, to restrict their
values.

Clearly, to solve this problem we have to take account of
certain new principles. It would be the best if we take
advantage of the principles which are no less general
than those used as the basis for the effective theory.

Following \cite{1}, we study here the possibility  to attract for
this purpose the suitably formulated old good principle of
maximal analyticity along with the polynomial boundedness
requirement for tree-level amplitude (first suggested in
\cite{2}; see also \cite{2a}). Some arguments (as well as the
corresponding list of references) in favor of the latter
requirement are given in \cite{1}. Thus we need to explain here
in more detail both the motivation and the exact formulation of
the maximal analyticity principle. The best way to do this is to
consider the simple example: the low energy elastic scattering
of two identical pseudoscalar particles with the mass
$\mu \ll m_1$, where $m_1$ stands for the mass of the lightest
allowed resonance. In this case the low energy effective
tree-level amplitude takes a form
\begin{equation}   \label{2.1}
A(s,t,u)=\sum_{i,j,k}^{}a_{ijk}s^i t^j u^k.
\end{equation}
Here summation in $i,j,k$ is infinite by the very meaning of the
term "effective", $a_{ijk}$ is completely symmetric in its
indices and the problem of LEC's is precisely that of $a_{ijk}$.
To the first glance these constants seem to be free parameters of
our effective theory. However, this -- widely believed -- point
of view is not quite correct. Below we demonstrate that certain
limitations on the values of $a_{ijk}$ follow directly from the
natural requirements of analyticity.

Here it is pertinent to recall one of the basic principles of
$S$-matrix theory, namely, the maximal analyticity principle. It
says, that the only singularities of a given process amplitude
are those required by the unitarity relation. In the framework of
field theory approach this relation is realized perturbatively
via the loop expansion scheme. This scheme automatically
generates all \underline{necessary} singularities required by
unitarity. Besides, it might develop also the
\underline{unnecessary} singular structures if the corresponding
terms are contained in the tree-level amplitudes.

Thus we conclude that, to avoid a contradiction with the maximal
analyticity principle, one has to take the tree-level amplitudes
as regular as possible.

Clearly, the singular structure of tree-level amplitudes is a
matter of model. The effective amplitude of the elastic
photon-photon scattering provides an example when this structure
is very complicated. However, it is well known that this feature
is uniquely connected with the existence of electron which --
together with photon -- has to be taken into account when
constructing the full system of states in QED. If the electron
field is included in Lagrangian as a separate degree of freedom,
the analytical structure of the lowest order amplitudes becomes
simple, the photon-photon scattering appearing as one of the
higher-order processes.

Extremely interesting analysis of the similar effect in a
framework of "toy-theory" --- the quark-level linear
$\sigma$-model --- has been implemented in the recent paper
\cite{6a}. The authors show that the double counting problem
(appearing due to dynamically generated additional scale) can be
solved using the compositeness condition (see Chapter 10 in
\cite{6}), the result providing a natural self-consistent
field-theoretic interpretation in terms of {\it either}
elementary particle {\it or} the bound state.

Now, let us come back to the tree-level amplitude (\ref{2.1}).
To fix its singular structure we can rely on the hypothesis of
quark-hadron duality which says that the full set of colorless
quark-gluon states is equal to the full set of hadronic states.
This can be also formulated as follows: the functional integral
for $S$-matrix in QCD can be identically rewritten in terms of
hadronic fields. Leaning upon this statement one concludes that
the singular structure of tree-level amplitude (\ref{2.1}) is
completely determined by the contributions of relevant
one-particle hadronic states. In other words, the quark-hadron
duality together with maximal analyticity principle require of
the tree-level amplitude of a given binary process to be a
meromorphic function of 3 (dependent) Mandelstam variables, the
only allowed singularities being just simple poles and the
ambiguity points (see Sec.~3 below).

The above reasoning allows us to avoid the refereeing to the
large-$N_c$ limit of QCD (cf. with \cite{7}). Moreover, it allows
one to reduce the very difficult (from the purely
phenomenological point of view) problem of LEC's to the problem
of spectrum parameters -- on-shell triple couplings and masses
(see \cite{1}).

\section{Cauchy's form in the case of two variables}
\mbox{}

The main tool used in \cite{1} to carry out the analytic
continuation, connecting the direct- and cross-chanel tree-level
amplitudes, is based on the Mittag-Leffler theorem in its
constructive form provided by the Cauchy method. This method
allows one to write down a general expression (which we call
below as Cauchy's form) for the polynomially bounded meromorphic
function $f(z)$ of one complex variable $z$, with given poles
$p_n$  $(n=1,...)$, corresponding principle parts $g_n(z)$ and
the degree $N$ of bounding polynomial. This expression reads

\begin{equation}    \label{3.1}
f(z)=\sum_{n=0}^{N}\frac{1}{n!}f^{(n)}(0)z^n +
\sum_{p=1}^{\infty}[g_p(z)-h_p^{(N)}(z)].
\end{equation}
Here
$$
h_p^{(N)}(z)\equiv \sum_{n=0}^{N}\frac{g_p^{(n)}(0)}{n!} z^n ,
$$
are the so-called correcting polynomials needed to ensure the
convergence of the infinite sum of pole contributions. It is
implied that $f(0)$ is regular, otherwise, the corresponding
principal part $g_0(z)$ should be added to the right hand side
of the Eq. (\ref{3.1}).

A rigorous proof of the form (\ref{3.1}) can be found in
textbooks on complex analysis (see, e.g., \cite{8,9}). However,
in our work we use the generalized version of (\ref{3.1})
allowing one to consider meromorphic functions of two complex
variables $(\nu , x)$. As far as we know, such a form could
hardly be found in the literature. Therefore, it makes sense to
give here a sketch of the proof of the generalized version of
(\ref{3.1}) most suitable for our needs. Later on we imply that
the reader is familiar with the case of one complex variable.

First of all we would like to remind the reader, that every
meromorphic function of two (and more) complex variables
$f(z_1, z_2)$ has two different kinds of singularities: poles
and the ambiguity points. The last term can be best explained by
the following example. Consider
$$
f(z_1, z_2) = \frac{z_1}{z_2}\ .
$$
This is a meromorphic function, its polar set being the
hyperplane $(z_1, 0)$ except the point $(0,0)$ which is precisely
the ambiguity point. The value of $f(z_1, z_2)$ at this point
depends of the path chosen to reach it. For example,
$$
\lim_{z_2 \rightarrow 0} \lim_{z_1 \rightarrow 0}
 f(z_1, z_2) = 0 ,
$$
while
$$
\lim_{z_1 \rightarrow 0} \lim_{z_2 \rightarrow 0}
 f(z_1, z_2) = \infty .
$$
Less trivial example is provided by Fig.1 where we show the
geography of the ambiguity points corresponding to the
string-like amplitude (\ref{4.1}) considered in Sec.~4 below.

In order to avoid unnecessary complications which have nothing to
do with the field-theoretical problems considered in our paper,
below (except the Sec.~4) we concentrate solely on a
consideration of the narrow class of meromorphic functions
$f(\nu, x)$ satisfying the following conditions:
\begin{itemize}
\item
They have only simple poles in each variable.
\item
They have no poles in both variables simultaneously.
\item
They have no fixed (i.e. independent of $x$) poles in $\nu$;
only moving poles of the form
\begin{equation}    \label{3.3}
\nu-x=Q_i\  ;\ \ \ \ \ \ \ \ \ \    \nu+x=-Q_i
\end{equation}
with
$$
0<Q_i<Q_{i+1}\ ,\ \ \ \ \ \ \ \ \     (i=1,2,...)
$$
are allowed.
\item
In the variable $x$ they have both moving poles of the form
(\ref{3.3}) and fixed poles at the points
$$
x=M_i^2\ ,\ \ \ \ \ \ \ \ \ \ \ \ \  (i=1,2,...),
$$
where
$$
0<M_i^2<M_{i+1}^2  .
$$
\end{itemize}

Nevertheless, it should be noted that the analysis of more
involved cases creates no difficulties.

To construct the generalized Cauchy form in $\nu$ (at fixed $x$)
we need to know asymptotics of $f(\nu,x)$ at large $\nu$. Since
it might depend on $x$, we have to consider the ratio
$$
\frac{f(\nu,x)}{\nu^{D(x)}}
$$
at large $\nu$. In the most interesting for us case, when the
Regge asymptotic condition is imposed (see \cite{1})
$$
D(x) = \alpha + \beta x\ .
$$
Let us introduce the step function
$$
N_x \equiv E[D(x)] + 1\ ,
$$
where $E[y]$ stands for the maximal integer less or equal to $y$.
This allows one to draw on the conventional definition of
polynomial boundedness (see, e.g., \cite{8,9}). Indeed, let us
consider real $x$ from a small interval $[a,b]$ such that
$$
N_a=N_x=N_b \equiv N\ .
$$
We say that the meromorphic function $f(\nu , x)$ is polynomially
bounded in $\nu$ in the band
$$
B_x\{\mid\nu\mid < \infty\ ,\ x\in [a,b]\}
$$
if there is a finite integer $N$ and infinite system of smooth
contours $C_p$ (circles with the radii
$R_{p+1}>R_p\ ,\ p=1,2,...$) in the  complex-$\nu$ plane such
that
\begin{equation}   \label{3.12}
\max_{x\in[a,b];\ \nu\in C_p} \left|
\frac{f(\nu ,x)}{\nu^{N+1}}\right|
\equiv M_p\stackrel{p\rightarrow\infty}{\longrightarrow 0}.
\end{equation}
The \underline{minimal} $N$ providing the correctness of the
uniform (in $x$) estimate (\ref{3.12}) we call as the degree of
bounding polynomial in $B_x$. This definition is equally applied
for both increasing and decreasing asymptotics, $N$ taking the
{\it negative} values in the latter case. The value $N\leq -1$
corresponds to the superconvergent asymptotic behavior.

It is important to stress that we consider the radii $R_p$ to be
independent of $x$. With the above definition in hand we can
immediately write down the generalized Cauchy form for the
meromorphic function $f(\nu, x)$, polynomially bounded (with the
degree $N$) in the band $B_x$. It looks as follows
\begin{equation}   \label{3.13}
f(\nu, x)=\sum_{k=0}^{N} \frac{1}{k!}
\frac{\partial^{(k)}f(0,x)}{\partial\nu^k} \nu^k +
\sum_{m=1}^{} \left[\frac{r_m(x)}{\nu-p_m(x)} -
h_m^{(N)}(\nu,x) \right]\ ,
\end{equation}
where
$$
h_m^{(N)}(\nu,x) = -\frac{r_m(x)}{p_m(x)}
\sum_{k=0}^{N} \frac{\nu^k}{p_m^k(x)}\ .
$$

The proof is based on a consideration of the following
contour integral in the \\ complex-$\nu$ plane:
$$
I_p(\nu, x,N) = \frac{1}{2\pi i}
\int\limits_{C_p}^{}\frac{\nu^{N+1}}{z^{N+1}}
\frac{f(z,x)}{(z-\nu)}dz\ .
$$
Exactly as in the case of one variable, it can be shown that at
every fixed $x\in [a,b]$
\begin{equation}    \label{3.16}
f(\nu, x)=\sum_{k=0}^{N} \frac{1}{k!}
\frac{\partial^{(k)}f(0,x)}{\partial\nu^k} \nu^k +
\sum_{m=1}^{p} \left[\frac{r_m(x)}{\nu-p_m(x)} -
h_m^{(N)}(\nu,x) \right] + I_p(\nu , x, N)\ .
\end{equation}
Taking the limit $p\rightarrow \infty $ and using the condition
(\ref{3.12}) one derives from (\ref{3.16}) the desired form
(\ref{3.13}) expressing the function $f(\nu ,x)$ of one complex
($\nu $) and one real ($x$) variable in the band $B_x$ as an
expansion in its poles in $\nu $.

>From the given above sketch one can derive the following
conclusions:
\begin{enumerate}
\item
Each item of the infinite sum in (\ref{3.13}) combines the
contributions from all the poles confined between $C_{m-1}$ and
$C_m$, i.e. from those with
$$
R_{m-1}<\mid p_m(x)\mid<R_m\ .
$$
Otherwise, the convergence of the summation procedure is
not guaranteed.
\item
At any $x\in[a,b]$ the partial fraction expansion (\ref{3.13})
converges uniformly everywhere in the complex-$\nu$ plane
except the small open vicinities of poles.
\item
At any fixed $\nu$ the form (\ref{3.13}) in the band $B_x$ can be
equally treated as the uniformly (in $x$) convergent series of
(analytic) functions of $x$. This property makes the expansion
(\ref{3.13}) a useful tool to carry out the analytic continuation
in $x$.
\item
The form (\ref{3.13}) certainly remains valid if the minimal
degree $N$ is changed for any integer $\tilde{N}>N$.
In this case, however, each of the series
\begin{equation}   \label{3.18}
S_n\equiv \sum_{m}^{}\frac{r_m(x)}{p_m^n(x)}
\end{equation}
with $n=N+1,...,\tilde{N}$ converges and, hence, it can
be summed independently. It is easy to show that
\begin{equation}    \label{3.19}
S_n + \frac{1}{n!}\left.
      \frac{\partial^{n}f(\nu,x)}{\partial\nu^n}
\right|_{\nu_{}^{} = 0} = 0\ ,\ \ \ \ (n=N+1,...,\tilde{N})\ .
\end{equation}
This, in turn, means that each unnecessarily high degree of $\nu$
taken into account in the correcting polynomials
$h_m^{(\tilde{N})}(\nu,x)$ is effectively cancelled by the
corresponding item appearing simultaneously in the first --
regular in $\nu$ -- term of (\ref{3.13}). Thus we conclude that
the Cauchy form (\ref{3.13}) presents a well defined rigid
construction allowing no twofold interpretation.
\item
With the properly chosen functions $r_i(x)$ and $p_i(x)$ one can
secure the convergence of the series (\ref{3.18}) even for $n<N$.
However, this does not mean that the equality (\ref{3.19}) is
also valid for $n<N+1$. The minimality of the declared degree
$N$ of a bounding polynomial corresponds to the necessary
presence of $\nu^N$ either in the regular term in (\ref{3.13})
or in correcting polynomials or in both terms simultaneously. In
other words, the \underline{actual} presence of $\nu^N$ in
(\ref{3.13}) mirrors the asymptotic behavior of the type
\begin{equation}   \label{3.20}
f(\nu ,x) \sim |\nu |^{D(x)}
\end{equation}
with $N \leq D(x) < N+1$ .

There is an important exception to the above formulated rule,
this exception being connected solely with our special choice
of the system of contours $C_p$ which we took symmetric with
respect to the origin of the complex-$\nu $ plane. This choice
results in a particular method of summation in (\ref{3.13}):
each item of the sum over poles contains the contributions of
all the poles with the same value of $|p_i|$. In the case when
$f(\nu ,x)$ is odd (even) in $\nu $, the correcting
polynomials are also odd (even). The same is true with respect
to the regular term. Thus, in this particular case the correcting
polynomials of the degree $N$ ensure the convergence of the
partial fraction expansion (\ref{3.13}) for the asymptotic low
(\ref{3.20}) with $D<N+2$, the presence of $\nu^N$ corresponding
to the asymptotic behavior (\ref{3.20}) with $N \leq D < N+2$\ .
\item
If the uniform in $x$ estimate (\ref{3.12}) is valid for
$x \in [a,b]$ with $N_a \not= N_b$ one can use the Cauchy form
(\ref{3.13}) with $N = \max\{N_a , N_b\}$ to present the function
$f(\nu ,x)$ in the band $B_x$.
\end{enumerate}

The usefulness of the technique developed in this Section is
explained by the fact that meromorphic functions with fiberwise
given Regge asymptotics appear naturally in the framework of
effective hadron field theory. This very technique (first
suggested in \cite{1}) is used throughout the remaining part of
our paper.

\section{String amplitudes and background interactions}
\mbox{}

The results of the previous section allow us to argue that:
\begin{description}
\item[A]
\underline{Every}
polynomially bounded meromorphic function $f(\nu,x)$ can be
presented in a form of the convergent series over its poles in
$\nu$ at those values of $x$ which correspond to the decreasing
asymptotic behavior in $\nu$ (we mean the contour asymptotics).
\item[B]
On the contrary, at those $x$ which correspond to a constant or
increasing asymptotics in $\nu$, \underline{none} of such
functions admit a representation constructed solely of the pole
contributions: the appearance of ~"background" terms
(polynomials in $\nu$ with coefficients depending on $x$) is
inevitable in this case.
\end{description}
In the light of these statements it is extremely instructive to
analyze the structure of the famous Veneziano ansatz \cite{10}
based on $B$-functions which are widely believed to be
constructed solely from resonances (for the review see
\cite{11}). By way of illustration we consider the simplest dual
(or, the same, string) amplitude without tachyon:
\begin{equation}   \label{4.1}
A(s,t) = [1- \alpha_1(s)- \alpha_1(t)]\
B\{1 - \alpha_1(s);\ 1 - \alpha_1(t)\}\ .
\end{equation}
It is implied that kinematical variables are chosen such that
$$\alpha_1(x)=\frac{1}{2} + x\ .$$
\begin{figure}  \label{1f}
\begin{center}
\begin{picture}(200,200)(0,0)

\put(0,100){\vector(1,0){200}}
\put(100,0){\vector(0,1){200}}

\multiput(40,100)(20,0){7}{\line(0,1){3}}
\multiput(40,100)(20,0){7}{\line(0,-1){3}}

\multiput(50,100)(20,0){6}{\line(0,1){1.5}}
\multiput(50,100)(20,0){6}{\line(0,-1){1.5}}

\multiput(100,40)(0,20){7}{\line(1,0){3}}
\multiput(100,40)(0,20){7}{\line(-1,0){3}}

\multiput(100,50)(0,20){6}{\line(1,0){1.5}}
\multiput(100,50)(0,20){6}{\line(-1,0){1.5}}

\put(32,105){\shortstack{ \mbox{ -3} } }
\put(155,105){\shortstack{ \mbox{ 3} } }

\put(105,38){\shortstack{ \mbox{ -3} } }
\put(105,158){\shortstack{ \mbox{ 3} } }

\put(110,190){\shortstack{ t } }
\put(190,90){\shortstack{ s } }

\multiput(50,150)(20,0){6}{\circle{3}}
\multiput(70,130)(20,0){5}{\circle{3}}
\multiput(90,110)(20,0){4}{\circle{3}}

\multiput(150,50)(0,20){6}{\line(0,1){2}}
\multiput(130,70)(0,20){5}{\line(0,1){2}}
\multiput(110,90)(0,20){4}{\line(0,1){2}}

\multiput(150,50)(0,20){6}{\line(1,0){2}}
\multiput(130,70)(0,20){5}{\line(1,0){2}}
\multiput(110,90)(0,20){4}{\line(1,0){2}}

\multiput(150,50)(0,20){6}{\line(0,-1){2}}
\multiput(130,70)(0,20){5}{\line(0,-1){2}}
\multiput(110,90)(0,20){4}{\line(0,-1){2}}

\multiput(150,50)(0,20){6}{\line(-1,0){2}}
\multiput(130,70)(0,20){5}{\line(-1,0){2}}
\multiput(110,90)(0,20){4}{\line(-1,0){2}}

\multiput(150,50)(0,20){6}{\circle*{1}}
\multiput(130,70)(0,20){5}{\circle*{1}}
\multiput(110,90)(0,20){4}{\circle*{1}}

\end{picture}
\end{center}

\caption{String-like amplitude
(\ref{4.1}): locations of ambiguity points.
\( \circ \) --- first series,
\mbox{\( + \) --- second} series,
\( \oplus \) --- superposition of two series
                 (so-called "Odorico zeros"). }

\end{figure}
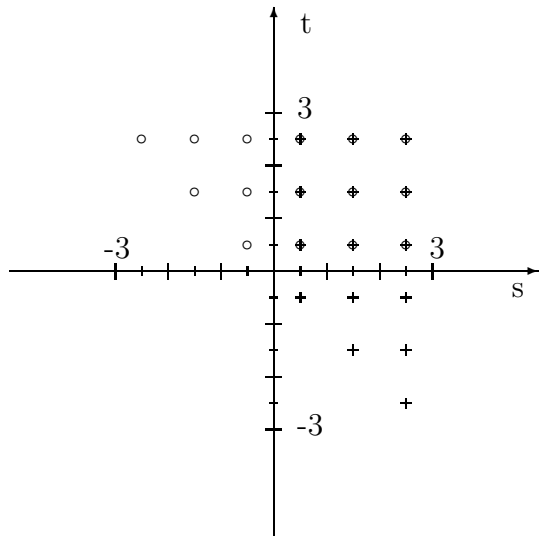
Fig.~1 shows the disposition of ambiguity points of $A(s,t)$;
note, that it reflects the space fibering structure
corresponding to the asymptotic behavior.

Since the pomeron contribution is not an issue here, one can
consider $A(s,t)$ as the amplitude of the process
$\pi^-\pi^+ \longrightarrow \pi^- \pi^+$. Let us study the
structure of $A(s,t)$ at $t=0$. In this case the actual (radial)
asymptotic behavior for $\arg{s}\neq0$ follows the Regge low
$A(s,0)\ \sim\ s^{1/2}$ which is also true with respect to
behavior on the system of circles $C_n$ with $R_n=n$. Hence,
according to the results of Sec.~3, the meaningful partial
fraction expansion for $A(s,0)$ cannot be written without
introducing of background terms of zeroth order in $\nu$. Let us
check this point. For this we need to know the principal parts
$g_n(s,0)$ at the poles
\begin{equation}   \label{4.4}
p_n=n + \frac{1}{2}\ ,\ \ \ \ \ \ n=0,1,...\ .
\end{equation}
Rewriting (\ref{4.1}) in the form
$$
A(s,0) = \sqrt{\pi}\,
\frac{\Gamma(\frac{1}{2}-s)}{\Gamma(-s)}
$$
and using the well-known formulae for $\Gamma$-function, we
obtain
\begin{equation}   \label{4.6}
g_n(s,0) = \frac{r_n(0)}{s-(n+\frac{1}{2})}\ ,
\end{equation}
where
\begin{equation}   \label{4.7}
r_n(0) = \frac{(2n+1)!!}{n!\ 2^{n+1}}\ .
\end{equation}
>From (\ref{4.6}) and (\ref{4.7}) it follows immediately that the
formal sum of principal parts ("nothing else but poles!")
$$\sum_{n=0}^{\infty}g_n(s,0)$$
diverges at every point of the complex-$s$ plane. Thus we
conclude that, in complete accordance with the statement $B$
above, the dual amplitude (\ref{4.1})
at $t=0$ along with the singular part (the sum of pole
contributions) contains also certain regular terms. It is not
difficult to write down the corresponding convergent
expansion. Bearing in mind that $A(0,0)$=0, we have (see Sec.~3):
\begin{equation}    \label{4.8}
A(s,0) = \sum_{n=0}^{\infty}
\left[\frac{r_n(0)}{s-(n+\frac{1}{2})} -
\frac{r_n(0)}{(n+\frac{1}{2})}\right].
\end{equation}
It can be easily shown that the series (\ref{4.8}) converges
uniformly and absolutely everywhere in the complex-$s$ plane
except small open vicinities of poles. This follows from
the absolute convergence of the series
$$\sum_{n=0}^{\infty}\frac{r_n(o)}{(n+\frac{1}{2})^2}
 \equiv\sum_{n=0}^{\infty}a_n\ ;$$
the latter, in turn, can be shown with the help of Gauss's test
$$\left|\frac{a_{n+1}}{a_n}\right| =
1 + \frac{A}{n} + O(\frac{1}{n^2})\ ,$$
because in our case $A=-3/2<-1$.

The similar analysis for A(s,-1) shows that the series of
principal parts
\begin{equation}   \label{4.9}
A(s,-1) = \sum_{n=o}^{\infty}
\frac{r_n(-1)}{s - (n+\frac{1}{2})}\ ,
\end{equation}
where
$$
r_n(-1) = -\frac{(2n-1)!!}{n!\ 2^{n+1}}\ ,
$$
converges by itself (also uniformly and absolutely) and, hence,
there is no necessity to take account of any background terms
associated with the correcting polynomials. This is precisely
the result which one would expect in accordance with the known
asymptotic behavior $A(s,-1)\sim s^{-\frac{1}{2}}\ .$
It provides an illustration to the statement $A$.

To fill a gap between $t=0$ and $t=-1$ in the above reasoning we
need to account for the explicit dependence of $r_n(t)$ and
$A(0,t)$ of the variable $t$. In the case under consideration
this can be easily done because we can take advantage of the
relation
$$
\frac{\Gamma(z) \Gamma(a+1)}{\Gamma(z+a)}=
\sum_{n=0}^{\infty}\frac{(-1)^n}{n!}a(a-1)(a-2)\ldots (a-n)
\frac{1}{z+n}\ ,
$$
which is valid for $a>0$.
Transforming the left hand side
$$
\frac{\Gamma (z)\Gamma (a+1)}{\Gamma (z+a)}=
\frac{(z+a)}{(a+1)}\ \frac{\Gamma (z)\Gamma (a+2)}
{\Gamma (z+a+1)}
$$
and taking
$$
a=-\frac{1}{2}-t\ ,\ \ \ \ \ \ \
z=\frac{1}{2}-s\ ,
$$
one obtains
\begin{equation}   \label{4.14}
A(s,t)=\sum_{n=0}^{\infty}\left[\frac{r_n(t)}{s-p_n}+
\frac{r_n(t)}{p_n}+u_n(t)\right]\ .
\end{equation}
Here $p_n$ are defined in (\ref{4.4}),
$$
r_n(t)=\frac{1}{n!}(\frac{1}{2}+t)(\frac{3}{2}+t)\ldots
[(n+\frac{1}{2})+t]\ ,
$$
and
$$
u_0(t)=-2t\ ; \ \ \ \ \ \
u_n(t)=-\frac{t}{n!} \frac{(\frac{1}{2}+t)(\frac{3}{2}+t)
\ldots [(n-\frac{1}{2})+t]}{(n+\frac{1}{2})}\ , \ \ \ \ \ \
(n=1,2,\ldots )\ .
$$
The expression (\ref{4.14}) applies for
\begin{equation}    \label{4.17}
t<+\frac{1}{2}\ .
\end{equation}
To put it into Cauchy's form we note that the series
\begin{equation}   \label{4.18}
\sum_{n=0}^{\infty}u_n(t)
\end{equation}
under the condition (\ref{4.17}) converges by itself and, hence,
can be summed independently. The resulting expression for the
amplitude $A(s,t)$ reads
\begin{equation}   \label{4.19}
A(s,t)=\frac{\Gamma (\frac{1}{2}) \Gamma (\frac{1}{2}-t)}
{\Gamma(-t)}+\sum_{n=0}^{\infty}
\left[\frac{r_n(t)}{s-p_n}+\frac{r_n(t)}{p_n}\right]\ .
\end{equation}
This is precisely the desired Cauchy's form valid for
$t<+\frac{1}{2}$. It is easy to check that at $t=0$ the
expression (\ref{4.19}) consisted with (\ref{4.9}). In contrast
with (\ref{4.18}), the series of correcting polynomials
\begin{equation}    \label{4.20}
\sum_{n=0}^{\infty}h_{n}^{(0)}(t)=
\sum_{n=0}^{\infty}\frac{r_n(t)}{p_n}
\end{equation}
diverges at $t>-\frac{1}{2}$ and could not be summed separately.

Now we can trace in more detail what happens with (\ref{4.19})
when $t$ crosses the boundary value $t=-\frac{1}{2}$
corresponding to the change of the asymptotic regime
$A(s,t)\sim s^{\alpha_1 (t)}$ from $\alpha_1 (t)\geq 0$ to
$\alpha_1 (t)< 0$. It can be easily shown that at
$t<-\frac{1}{2}$ the series (\ref{4.20}) converges, the summation
giving the result (cf. with (\ref{3.19}))
$$
\sum_{n=0}^{\infty}h_n^{(0)}(t)=
-\frac{\Gamma (\frac{1}{2})\Gamma (\frac{1}{2}-t)}
{\Gamma(-t)}\ , \ \ \ \ \ \ \
(t<-\frac{1}{2})\ .
$$

Thus, in complete accordance with the general scheme discussed
above (see Sec.~3), we conclude that at $t<-\frac{1}{2}$ the
amplitude (\ref{4.1}) admits a representation
\begin{equation}     \label{4.22}
A(s,t)= \sum_{n=0}^{\infty} \frac{r_n(t)}{s-p_n}
\end{equation}
constructed solely from the resonance contributions; at $t=-1$
this form coincides with (\ref{4.9}). The expression (\ref{4.22})
is oftenly used in the literature to show the physical content of
the string amplitude (\ref{4.1}).

The important conclusion to be drawn from the above analysis is
that the conventional dual (string) models of hadrons are based
on three rather different general postulates, to say nothing
about suggestions of a particular nature. First, they take
advantage of the crossing symmetry requirement. Second, they are
rested on certain analyticity conditions, namely, those of
meromorphy and polynomial boundedness. Third, they imply --
though in a highly latent form -- a particular suggestion about
the (unique!) connection between the direct channel spectrum
parameters and the point-like (background) interactions, this
connection explicitly revealing only in the band of the momentum
transfer $t$ corresponding to $\alpha_I(0)\geq0$.

The two first postulates are quite general, whereas the third one
is nothing but an artifact of the ansatz based on $B$-functions.
Thus it looks reasonable to consider a theory which is free of
any particular suggestions about the structure of point-like
vertices. This is precisely the way which we follow here.

\section{Bootstrap equations for the parameters  \\
of $(\pi,K)$ resonances}
\mbox{}

Let us now turn to a consideration of $(\pi,K)$ processes. Unlike
\cite{1}, here we are interested mostly in derivation of the
complete set of bootstrap constraints. For this we use the Cauchy
forms in the bands $B_s, B_t$ and $B_u$ corresponding to three
cross-conjugated channels. In contrast with \cite{1}, we write
those forms in terms of independent pairs of kinematical
variables $\{\nu_x, x\}\ \ (x=s,t,u)$ and impose more realistic
-- Regge -- asymptotic requirements. Since the logical scheme,
compared to that described in \cite{1}, remains unchanged, we
omit unnecessary comments. The summary of relevant formulae and
notations is given in Appendix (see also \cite{1}); Fig.~2
explains the geography of bands $B_x$ and domains $D_x$.

\begin{figure} \label{2f}
\begin{center}

\begin{picture}(200,200)(0,0)
\put(85,170){\vector(-1,-2){70}}
\put(10,60){\vector(1,0){140}}
\put(125,30){\vector(-1,2){75}}

\multiput(30,40)(1,2){60}{\circle*{0.2}}
\multiput(10,70)(2,0){60}{\circle*{0.2}}
\multiput(110,40)(-1,2){60}{\circle*{0.2}}

\multiput(10,40)(1,2){70}{\circle*{0.2}}
\multiput(10,50)(2,0){60}{\circle*{0.2}}
\multiput(130,40)(-1,2){70}{\circle*{0.2}}

\multiput(15,50)(10,20){2}{\line(1,0){20}}
\multiput(15,50)(1.5,0){14}{\line(1,2){10}}

\multiput(105,50)(-10,20){2}{\line(1,0){20}}
\multiput(105,50)(1.5,0){14}{\line(-1,2){10}}

\multiput(70,120)(-0.7,1.4){15}{\line(1,2){10}}
\multiput(70,120)(10,20){2}{\line(-1,2){10}}

\put(70,30){\shortstack{ $\underline{ B_t} $}}
\put(75,55){\line(0,-1){15}}

\put(15,115){\shortstack{ $\underline{ B_s} $}}
\put(45,105){\line(-2,1){15}}

\put(112,115){\shortstack{ $\underline{ B_u} $}}
\put(95,105){\line(2,1){15}}

\put(70,60){\vector(0,1){15}}
\put(72,73){\shortstack{ $ t $}}

\put(50,100){\vector(2,-1){12}}
\put(57,87){\shortstack{ $ s $}}

\put(90,100){\vector(-2,-1){12}}
\put(72,95){\shortstack{ $ u $}}

\put(70,60){\circle*{2}}
\put(90,100){\circle*{2}}
\put(50,100){\circle*{2}}

\put(25,65){\line(-1,2){10}}
\put(0,90){\shortstack{ $\underline{ D_u} $}}

\put(115,65){\line(1,2){10}}
\put(125,90){\shortstack{ $\underline{ D_s} $}}

\put(75,140){\line(2,1){20}}
\put(95,155){\shortstack{ $\underline{ D_t} $}}

\put(140,50){\shortstack{ $ \nu_t $}}
\put(40,170){\shortstack{ $ \nu_u $}}
\put(20,30){\shortstack{ $ \nu_s $}}

\end{picture}
\end{center}

\caption{Disposition of the bands
$B_x$ and intersection domains
$D_x \; (x = s , t , u) $.}
\end{figure}
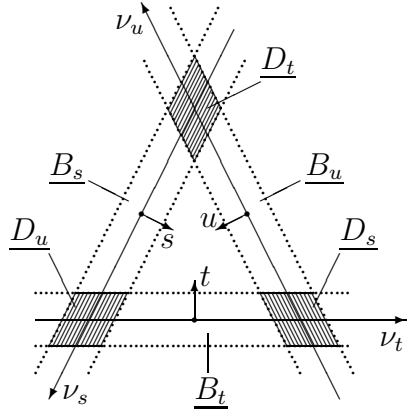


Let us begin our analysis from the band $B_s$. In this case
$\nu_s=(t-u)$ is considered as a complex variable, while $s$ --
as a small real parameter $(|s|\sim 0)$. So, the Cauchy form
(\ref{3.13}) for the combination $\left(A+2B\right)_{B_s}$
with the principal parts defined according to (\ref{A.13}) and
(\ref{A.14}) (under the condition (\ref{A.9}) fixing the bounding
polynomial degree $N=0$) reads
\begin{eqnarray}
\left(A+2B\right)_{B_s}=\alpha_s(s)+
2\sum_{(I=0)}^{}G_0P_J\left(\frac{\Sigma+2s}{4F}\right)
\left\{\frac{1}{\nu_s+(s+2\theta)}-\frac{1}{s+2\theta}\right\}\
\nonumber  \\
+4\sum_{(I=1)}^{}G_1P_J\left(\frac{\Sigma+2s}{4F}\right)
\left\{\frac{1}{\nu_s+(s+2\theta)}-\frac{1}{s+2\theta}\right\}\
\nonumber  \\
+2\sum_{(I=1/2)}^{}G_{1/2}P_J\left(1-\frac{\Sigma+s}{2\Phi}\right)
\left\{\frac{1}{\nu_s-(s+2\theta)}+\frac{1}{s+2\theta}\right\}.
\label{5.1}
\end{eqnarray}
Here $\alpha_s$ stands for unknown function of $s$.

The analogous form for $\left(A-B\right)_{B_s}$ reads
\begin{eqnarray}
\left(A-B\right)_{B_s}=
& \ & 2 \sum_{(I=0)}^{}G_0P_J\left(\frac{\Sigma+2s}{4F}\right)
\frac{1}{\nu_s+(s+2\theta)}\
\nonumber  \\
& - & 2 \sum_{(I=1)}^{}G_1P_J\left(\frac{\Sigma+2s}{4F}\right)
\frac{1}{\nu_s+(s+2\theta)}\
\nonumber  \\
&- & 4\sum_{(I=1/2)}^{}
               G_{1/2}P_J\left(1-\frac{\Sigma+s}{2\Phi}\right)
\frac{1}{\nu_s-(s+2\theta)}\ .
\label{5.2}
\end{eqnarray}
In the latter case no unspecified functions of $t$ appear in the
Cauchy form because -- according to (\ref{A.9}) -- the degree of
the relevant bounding polynomial is negative ($N=-1$). From
(\ref{5.1}) and (\ref{5.2}) one derives
\begin{eqnarray}
\left(A\right)_{B_s}=\frac{1}{3}\alpha_s(s)+
2\sum_{(I=0)}^{}G_0P_J\left(\frac{\Sigma+2s}{4F}\right)
\left\{\frac{1}{\nu_s+(s+2\theta)}-
\frac{1}{3}\ \frac{1}{s+2\theta}\right\}\
\nonumber  \\
-\frac{4}{3}\sum_{(I=1)}^{}
                G_1P_J\left(\frac{\Sigma+2s}{4F}\right)
\frac{1}{s+2\theta}\ \ \ \
\nonumber  \\
-2\sum_{(I=1/2)}^{}
        G_{1/2}P_J\left(1-\frac{\Sigma+s}{2\Phi}\right)
\left\{\frac{1}{\nu_s-(s+2\theta)}-
\frac{1}{3}\ \frac{1}{s+2\theta}\right\},
\label{5.3}
\end{eqnarray}
and
\begin{eqnarray}
\left(B\right)_{B_s} &=&\frac{1}{3}\alpha_s(s)-
\frac{2}{3}\sum_{(I=0)}^{}
                G_0P_J\left(\frac{\Sigma+2s}{4F}\right)
\frac{1}{s+2\theta}\ \ \ \
\nonumber  \\
&+& 2\sum_{(I=1)}^{}G_1P_J\left(\frac{\Sigma+2s}{4F}\right)
\left\{\frac{1}{\nu_s+(s+2\theta)}-
\frac{2}{3}\ \frac{1}{s+2\theta}\right\}\
\nonumber  \\
&+& 2\sum_{(I=1/2)}^{}
            G_{1/2}P_J\left(1-\frac{\Sigma+s}{2\Phi}\right)
\left\{\frac{1}{\nu_s-(s+2\theta)}+
\frac{1}{3}\ \frac{1}{s+2\theta}\right\}.
\label{5.4}
\end{eqnarray}

The similar consideration in the band $B_t$
results in the expressions
\begin{eqnarray}
\left(A\right)_{B_t}=a_0(t)-
2\!\!\!\!\sum_{(I=1/2)}^{}G_{1/2}P_J
\left(1+\frac{t}{2\Phi}\right)
\left\{\frac{1}{\nu_t-(t+2\theta)}-
\frac{1}{\nu_t+(t+2\theta)}+
\frac{2}{t+2\theta}\right\}\!\!,
\label{5.5}
\end{eqnarray}
\begin{eqnarray}
\left(B\right)_{B_t}=
-2\sum_{(I=1/2)}^{}G_{1/2}P_J\left(1+\frac{t}{2\Phi}\right)
\left\{\frac{1}{\nu_t-(t+2\theta)}+
\frac{1}{\nu_t+(t+2\theta)}\right\}.
\label{5.6}
\end{eqnarray}
Here $a_0(t)$ is another unknown function of $t$.

At last, in $B_u$ one has
\begin{eqnarray}
\left(A\right)_{B_u}  =&& \frac{1}{3}\alpha_u(u)-
2\sum_{(I=0)}^{}G_0P_J\left(\frac{\Sigma+2u}{4F}\right)
\left\{\frac{1}{\nu_u+(u+2\theta)}-
\frac{1}{3}\frac{1}{u+2\theta}\right\}
\nonumber  \\
&-& \frac{4}{3}\sum_{(I=1)}^{}
                G_1P_J\left(\frac{\Sigma+2u}{4F}\right)
\frac{1}{u+2\theta}
\nonumber  \\
&+& 2\sum_{(I=1/2)}^{}
        G_{1/2}P_J\left(1-\frac{\Sigma+u}{2\Phi}\right)
\left\{\frac{1}{\nu_u+(u+2\theta)}+
\frac{1}{3}\frac{1}{u+2\theta}\right\},
\label{5.7}
\end{eqnarray}
\begin{eqnarray}
\left(B\right)_{B_u}=&-&\frac{1}{3}\alpha_u(u)+
\frac{2}{3}\sum_{(I=0)}^{}
                G_0P_J\left(\frac{\Sigma+2u}{4F}\right)
\frac{1}{u+2\theta}
\nonumber  \\
&+&2\sum_{(I=1)}^{}G_1P_J\left(\frac{\Sigma+2u}{4F}\right)
\left\{\frac{1}{\nu_u-(u+2\theta)}+
\frac{2}{3}\frac{1}{u+2\theta}\right\}
\nonumber  \\
&+&2\sum_{(I=1/2)}^{}
           G_{1/2}P_J\left(1-\frac{\Sigma+u}{2\Phi}\right)
\left\{\frac{1}{\nu_u+(u+2\theta)}-
\frac{1}{3}\frac{1}{u+2\theta}\right\},
\label{5.8}
\end{eqnarray}
where $\alpha_u(u)$ is the third unknown function.

The system of relations (\ref{5.3}) --- (\ref{5.8}) exhausts
constructive potentialities of our postulates. The further
progress can be only based on the analysis of this system.

First, we need to specify the form of three unknown functions
$a_0(t), \alpha_s(s)$ and $\alpha_u(u)$ in terms of the
resonance spectrum parameters. Precisely as in \cite{1}, this
can be done through a comparison of the pairs of relevant forms
in the domains of mutual intersection of the corresponding bands.
Thus we have to analyze the following conditions $(X=A,B)$:
\begin{equation}    \label{5.9}
X\left(\left.M\right/_{B_t} \right)=
X\left(\left.M\right/_{B_u} \right)\ ,\ \ \ \ \ \
M\in D_s=B_t\cap B_u\ ;
\end{equation}
\begin{equation}    \label{5.10}
X\left(\left.M\right/_{B_u} \right)=
X\left(\left.M\right/_{B_s} \right)\ ,\ \ \ \ \ \
M\in D_t=B_s\cap B_s\ ;
\end{equation}
\begin{equation}   \label{5.11}
X\left(\left.M\right/_{B_s} \right)=
X\left(\left.M\right/_{B_t} \right)\ ,\ \ \ \ \ \
M\in D_u=B_s\cap B_t\ .
\end{equation}
The analysis is simple though tedious. Therefore, we give here
only general outline and show the final results.

To present our results in a compact form we introduce three
auxiliary functions depending on two real variables $x$ and
$\nu$. They are the following
\begin{eqnarray}
\Phi(x,\nu) & \stackrel{\rm def}{=} &
-\sum_{(I=0)}^{}G_0
\frac{P_J\left(\frac{\nu+x-M^2}{4F}\right)}{x-M^2}
-\sum_{(I=1)}^{}G_1
\frac{P_J\left(\frac{\nu+x-M^2}{4F}\right)}{x-M^2}
\nonumber \\
& &-4\sum_{(I=\frac{1}{2})}^{}G_{1/2}
\left[
\frac{P_J
         \left(1+\frac{\nu+x-2\theta}{4\Phi}\right)
    - P_J
         \left(1+\frac{x}{2\Phi}\right)}
     {\nu-(x+2\theta)}-
\frac{P_J\left(1+\frac{x}{2\Phi}\right)}{x+2\theta}
\right],
\label{5.12}
\end{eqnarray}
\begin{eqnarray}
\Psi_1(x,\nu)\stackrel{\rm def}{=}
3\left\{
\sum_{(I=0)}^{}G_0
  \left[
\frac{P_J\left(\frac{\Sigma+2x}{4F}\right)
    - P_J\left(\frac{\nu+x-M^2}{4F}\right)}
     {\nu-(x+2\theta)}+
\frac{2}{3}\frac{P_J\left(\frac{\Sigma+2x}{4F}\right)}
                {x+2\theta}
  \right]
  \right.\ \ \
\nonumber \\
  \left.
+\sum_{(I=1)}^{}G_1
  \left[
\frac{P_J\left(\frac{\Sigma+2x}{4F}\right)
    - P_J\left(\frac{\nu+x-M^2}{4F}\right)}
     {\nu-(x+2\theta)}+
\frac{4}{3}\frac{P_J\left(\frac{\Sigma+2x}{4F}\right)}
                {x+2\theta}
  \right]
  \right.\ \ \
\nonumber \\
  \left.
-\sum_{(I=\frac{1}{2})}^{}G_{1/2}
  \left[
\frac{P_J\left(1+\frac{\nu+x-2\theta}{4\Phi}\right)}
     {x-M^2}+
\frac{2}{3}\frac{P_J\left(1-\frac{\Sigma+x}{2\Phi}\right)}
                {x+2\theta}
  \right]
 \right\},
\label{5.13}
\end{eqnarray}
\begin{eqnarray}
\Psi_2(x,\nu)\stackrel{\rm def}{=}
 2\sum_{(I=0)}^{}G_0
\frac{P_J\left(\frac{\Sigma+2x}{4F}\right)}
                {x+2\theta}
-6\sum_{(I=1)}^{}G_1
P_J\left(\frac{\Sigma+2x}{4F}\right)
 \left[
\frac{1}{\nu+(x+2\theta)}-
\frac{2}{3}\frac{1}{x+2\theta}
 \right]\
\nonumber \\
-3\sum_{I=\frac{1}{2}}^{}G_{1/2}
  \left[
2\frac{P_J\left(1-\frac{\Sigma+x}{2\Phi}\right)-
       P_J\left(1-\frac{\nu+x-2\sigma}{4\Phi}\right)}
      {\nu-(x+2\theta)}
 \right.\ \ \ \ \ \
\nonumber \\
 \left.
 +\frac{2}{3}\frac{P_J\left(1-\frac{\Sigma+x}{2\Phi}\right)}
                   {x+2\theta}
+\frac{P_J\left(1-\frac{\nu+x-2\sigma}{4\Phi}\right)}
       {x-M^2}
  \right].
\label{5.14}
\end{eqnarray}
Each of the above constructions presents well-defined expression
near the point $(x=0, \\ \nu=2\sigma)$. This is not an
assumption: the expressions (\ref{5.12})--(\ref{5.14}) appear
naturally during the analysis of consistency requirements
(\ref{5.9})--(\ref{5.11}), and their convergency follows
directly from our asymptotic conditions.

Let us start the consideration from the domain $D_s$. The first
of the conditions (\ref{5.9}), namely,
$$
\left. B \right/_{B_t} = \left. B \right/_{B_u}
$$
gives
\begin{eqnarray}
\alpha_u(u)=\Psi_2(u,-\nu_u)\ , \ \ \ \ \
(u\sim 0,\  \nu_u \sim -2\sigma)\ .
\label{5.15}
\end{eqnarray}
The second independent condition (\ref{5.9})
$$
\left.(A+B) \right/_{B_t} = \left.(A+B)\right/_{B_u}
$$
results in the equality
\begin{eqnarray}
a_0(t)=\Phi(t,\nu_t),\ \ \ \ \ \
(t\sim 0,\  \nu_t \sim 2\sigma).
\label{5.16}
\end{eqnarray}
No other conclusions can be drawn from the condition (\ref{5.9})
and the expressions (\ref{5.5})--(\ref{5.8}).

Similarly, the analysis of the condition (\ref{5.10}) results in
the following expressions for $\alpha_s(s)$ and $\alpha_u(u)$
in $D_t$
\begin{eqnarray}
\alpha_s(s)=\Psi_1(s,-\nu_s)\ , \ \ \ \ \
(s\sim 0,\  \nu_s \sim -2\sigma)\ .
\label{5.17}
\end{eqnarray}
\begin{eqnarray}
\alpha_u(u)=\Psi_1(u,\nu_u)\ , \ \ \ \ \
(u\sim 0,\  \nu_u \sim 2\sigma)\ .
\label{5.18}
\end{eqnarray}
At last, from (\ref{5.11}) it follows that in $D_u$
\begin{eqnarray}
a_0(t)=\Phi(t,-\nu_t),\ \ \ \ \ \
(t\sim 0,\  \nu_t \sim -2\sigma).
\label{5.19}
\end{eqnarray}
\begin{eqnarray}
\alpha_s(s)=\Psi_2(s,\nu_s)\ , \ \ \ \ \
(s\sim 0,\  \nu_s \sim 2\sigma)\ .
\label{5.20}
\end{eqnarray}
The relations (\ref{5.15})--(\ref{5.20}) are the only
restrictions following from the compatibility conditions
(\ref{5.9})--(\ref{5.11}) for the amplitudes $A$ and $B$ defined
by the forms (\ref{5.3})--(\ref{5.4}) in $B_s$,
(\ref{5.5})--(\ref{5.6}) in $B_t$ and
(\ref{5.7})--(\ref{5.8}) in $B_u$.

To proceed further, it is convenient to separate the restrictions
(\ref{5.15})--(\ref{5.20}) into two independent groups. Noting,
that each of the functions $a_0(t),\  \alpha_s(s)$ and
$\alpha_u(u)$ depends only on {\it one} variable, we conclude
that the dependence of $\Phi(x,\nu_x),\  \Psi_1(x,\nu_x)$ and
$\Psi_2(x,\nu_x)$ on $\nu_x$ is purely fictitious. In other
words, to compute the left hand sides of (\ref{5.15})--
(\ref{5.20}) one can assign to $\nu_s,\  \nu_t,\  \nu_u$ any
arbitrary values from the validity domain $\nu_x \sim 2\sigma$
of the forms (\ref{5.12})--(\ref{5.14}).

The above note allows one to rewrite (\ref{5.15})--(\ref{5.20})
in the form of two independent systems. The first one reads
\begin{eqnarray}
a_0(t) & = & \Phi_{}(t,2\sigma)\ , \nonumber  \\
\alpha_s(s) & = & \Psi_1(s,2\sigma),
\label{5.21}   \\
\alpha_u(u) & = & \Psi_1(u,2\sigma).  \nonumber
\end{eqnarray}
It provides the desired explicit formulae expressing the
functions $a_0(t),\  \alpha_s(s)$ and $\alpha_u(u)$ in terms of
the resonance spectrum parameters: triple couplings $G_I$ and
masses $M^2, \mu^2, m^2$.

The second group consists of infinite set of self-consistency
conditions. It reads
\begin{eqnarray}
  \left. \frac{\partial^{k+p}\Psi_1(x,\nu)}
         {\partial x^k \partial \nu^p}\right/_Q
& = &  \left. \frac{\partial^{k+p}\Psi_2(x,\nu)}
         {\partial x^k \partial \nu^p}\right/_Q\ ;
\ \ \ \ \ \ \ \ \ \ \ \ \ \ \ \ \ \ \ \
\nonumber  \\
   \left. \frac{\partial^{k+p+1}\Psi_2(x,\nu)}
     {\partial x^k \partial \nu^{p+1}}\right/_Q & = & 0\ ;
\label{5.22}  \\
   \left. \frac{\partial^{k+p+1}\Phi(x,\nu)}
     {\partial x^k \partial \nu^{p+1}}\right/_Q & = & 0\ ;
\nonumber  \\
Q \equiv \{x=0,\  \nu = 2\sigma\}; &  & \ \ \ \ \ \ \ \
k,p = 0,1,2\ldots\ \ .
\nonumber
\end{eqnarray}
These conditions follow directly from the noted above
independence of $\Phi(x,\nu),\  \Psi_1(x,\nu)$ and
$\Psi_2(x,\nu)$ of the second argument and from the
equivalence of (\ref{5.17}) and (\ref{5.20}). Later on we call
this property as {\em reparameterization invariance}
(RP-invariance). The expressions (\ref{5.12})--(\ref{5.14})
provide the explicit forms of the {\em generating functions}
for the system (\ref{5.22}).

RP-invariance imposes very strong constraints on the values of
spectrum parameters. Those constraints mirror the general
properties of analyticity and crossing symmetry. In the
literature they are commonly called as \underline{bootstrap}.
\pagebreak

\section{Analysis of the bootstrap constraints}
\mbox{}

It would be the best if we could show the closed solution of the
system (\ref{5.22}). Unfortunately, we cannot. This is not only
connected with our inability to solve this infinite system, but
also with the obvious necessity to get a deeper understanding of
the form of each individual equation. Therefore, below we
concentrate mostly on the detailed semiphenomenological analysis
of few "lowest" equations corresponding to $k,p\leq 1$ in
(\ref{5.22}).

However, before starting this analysis we would like to point out
one important feature of the system (\ref{5.22}). Using the
expressions (\ref{5.12})--(\ref{5.14}) for the generating
functions $\Phi, \Psi_1$ and $\Psi_2$ one can easily show that
the scalar particles (both with $I=0$ and $I=1/2$) do not
contribute to this system at all. Thus no limitations on their
couplings and masses follow from (\ref{5.22}). This feature is
uniquely connected with the local character of our consideration.
We do not use any assumptions about the asymptotic behavior of
the amplitude outside the narrow band corresponding to the
momentum transfer close to zero. Compared to \cite{1}, we use
here much weaker (Regge) asymptotic restrictions which do not
require the decreasing behavior of the amplitudes of nonexotic
channels. This very reason explains the difference between the
corresponding systems of bootstrap constraints.

The above note clearly shows that -- as for now -- it would be
premature to seek the general solution of the system
(\ref{5.22}); additional information concerning the asymptotic
behavior at larger values of the momentum transfer should be
first taken into account.

Let us now turn to a consideration of the lowest order
bootstrap constraints (\ref{5.22}). Those constraints look
too bulky to be shown here explicitly. Therefore, below we show
their form in $SU_3$ chiral limit $m^2=\mu^2=0$. The only
exception will be made for the relation
\begin{equation}   \label{6.1}
\Psi_1(0,2\sigma)=\Psi_2(0,2\sigma)
\end{equation}
from the first group of the Eqs. (\ref{5.22}). With the help of
(\ref{5.12}) and (\ref{5.14}) it can be written in the form
\begin{equation}
\sum_{(I=1)}^{}\sum_{J=1,3,\ldots}^{}
\frac{G_1}{M^2\Sigma}P_J\left(\frac{\Sigma}{4F}\right)=
\sum_{(I=\frac{1}{2})}^{}\sum_{J=1,2,\ldots}^{}
\frac{G_{1/2}}{M^2\Sigma}
\left[1-P_J\left(1-\frac{\Sigma}{2\Phi} \right) \right]\ ,
\label{6.2}
\end{equation}
which is quite suitable for the numerical testing.

By construction (see Sec.~3) we have to carry out the summation
in (\ref{6.2}) in order of increasing mass. Thus the
contributions of the lightest mesons $(\rho$ and $K^{*})$ can be
separated without breaking the convergence of the remaining
series (such a trick with respect to the $lowest\  spin\  J=1$
contributions would be a mistake!). This gives
\begin{equation}   \label{6.3}
\frac{G_{\rho}}{M^2_{\rho}F_{\rho}}+\ldots =
\frac{2G_{K^{*}}}{M^2_{K^{*}}\Phi_{K^{*}}}+\ldots\ ,
\end{equation}
where ellipsis stand for the contribution due to heavier
($M\geq1.4~GeV$) mesons. Using the experimental data \cite{12}
and the expressions for $F$ and $\Phi$ (see Appendix) along with
the $SU_3$ estimate for $G_{\rho}$
\begin{equation}    \label{6.4}
\frac{G_{\rho}}{F_{\rho}}=\frac{2G_{K^{*}}}{\Phi_{K^{*}}}\ ,
\end{equation}
one obtains (in $GeV^{-2}$)
$$
(32\pm 2)+ \ldots = (24\pm 1.2)+ \ldots\ .
$$
The agreement does not look satisfactory. The reason becomes
clear when we take account of the contributions due to other
relatively light mesons listed in \cite{12} (with $M\leq 2~GeV$).
In this case we obtain the relation (see Appendix, Table 1)
$$
(34.5\pm 2.5)+ \ldots = (29.5\pm 2.5)+ \ldots\ ,
$$
which looks much more impressive. It is easy to understand that
the contributions of heavy mesons ($M\geq 2~GeV$) cannot destroy
the agreement. Indeed, for those mesons $2\sigma \ll M^2$ and,
hence, we can use the limit $\sigma =0$ when computing the
corresponding terms. In this case
$$
P_J\left(\frac{\Sigma}{4F}\right)=P_J(1)=1\ , \ \ \ \ \ \ \
P_J\left(1-\frac{\Sigma}{2\Phi}\right)=P_J(-1)=(-1)^J\ ,
$$
and we can rewrite (\ref{6.2}) as follows
\begin{eqnarray}
(34.5\pm 2.5)+\sum_{(I=1)}^{}\sum_{J=1,3,\ldots}^{}
\left.\frac{G_1}{M^4}
\right/_{M^2>2~GeV}=\ \
\nonumber \\
(29.5\pm 2.5)+\sum_{(I=\frac{1}{2})}^{}\sum_{J=1,3,\ldots}^{}
\left.\frac{2G_{1/2}}{M^4}\right/_{M^2>2~GeV}\ ,
\label{6.5}
\end{eqnarray}
where the sum in the right hand side does not contain
contributions of mesons with even spins at all. The (approximate)
equivalence of the sums in (\ref{6.5}) follows directly from
$SU_3$ symmetry.

The above analysis shows that the relative magnitude of two most
significant contributions --- those of $\rho$- and $K^*$-mesons
--- appearing in the lowest order bootstrap condition (\ref{6.1})
(based on the Regge asymptotic requirements), proves to be quite
consistent with the well established experimental data. This
conclusion remains also true with respect to the constraints of
higher orders. In particular, the corresponding relation of the
next --- compared to (\ref{6.2}) --- order follows from the
second group of bootstrap constraints (\ref{5.22}) at
$k=1,\  p=0$. In the limit $m^2=\mu^2=0$ it reads
\begin{equation}   \label{6.6}
\sum_{I=1}^{}\frac{G_1}{M_1^6}[J(J+1)-1] =
\sum_{I=1/2}^{}\frac{G_{1/2}}{M^6}[1-(-1)^J][J(J+1)-1]\ ,
\end{equation}
the correctness of the relationship among the contributions of
mesons with $I=1/2$ and $I=1$ being obvious. However the
numerical test of the SR (\ref{6.6}) is of less interest compared
to that of (\ref{6.2}) since in this case the contributions of
mesons with $J\geq2$ prove to be relatively more important.

Among the constrains (\ref{5.22}) there are also sum rules
containing the contributions of isoscalar mesons with
$J=2,4,...$ . For example, the relation
\begin{eqnarray}
-\sum_{I=0}^{}\frac{G_0}{M^4}J(J+1)+
\sum_{I=1}^{} \frac{G_1}{M^4}J(J+1) =
\sum_{I=1/2}^{}\frac{2G_{1/2}}{M^4} \left\{[1-(-1)^J] +
(-1)^JJ(J+1) \right\}
\label{6.7}
\end{eqnarray}
can be derived either from the first group of (\ref{5.22})
at $k+p=1$ or from the third one at $k+p=0$.

Unfortunately, as in the case of Eq. (\ref{6.6}), the existing
data on $\pi K$ resonances are not sufficient for the reliable
numerical analysis of (\ref{6.7}) .

The system of bootstrap constraints (\ref{5.22}) is based on the
Regge asymptotic requirements. It differs from the analogous
system following from the much stronger asymptotic conditions
considered in \cite{1}. This difference, in turn, corresponds to
different forms of the function $a_0(t)$ appearing in the
expression (\ref{5.5}) for the amplitude of elastic $\pi K$
scattering. To compute chiral coefficients we have to make a
choice between two possibilities. The phenomenological analysis
provides arguments in favor of the Regge-like bootstrap.
Therefore, when calculating LEC's in the next Section we use the
expression (\ref{5.16}) instead of that given in \cite{1}.

\section{Low energy coefficients, chiral duality \\
         and light scalars}
\mbox{}

Now we have all necessary ingredients to express the low energy
coefficients of the elastic scattering amplitudes $A(\nu_t,t)$
and $B(\nu_t,t)$ in terms of the spectrum parameters
$G_I,\  M^2,\  m^2$ and $\mu^2$. For this we need the expressions
(\ref{5.5}) and (\ref{5.6}) along with the bootstrap requirements
(\ref{5.16}) and (\ref{5.22}) allowing one to fix the form of
$a_0(t)$. To simplify the form of $A(\nu_t , t)$ it is
appropriate to choose the parameterization $\nu = 2\sigma - t$ in
(\ref{5.16}), this choice being permissible since we are only
interested in the values of $t$ close to zero.

The resulting Cauchy forms for $A(\nu_t, t)$ and $B(\nu_t, t)$
\begin{eqnarray}
A(\nu, t) = & - & \sum^{}_{(I=0)}G_0
\frac{P_J\left(\frac{\Sigma}{4F}\right)}{t-M^2_0}
           +\sum^{}_{(I=1)}G_1
\frac{P_J\left(\frac{\Sigma}{4F}\right)}{t-M^2_1}
\ \ \ \ \ \ \ \ \ \ \ \ \ \ \ \ \ \ \ \ \ \ \ \ \ \ \ \ \ \ \ \
\nonumber \\
        & + & 2\sum^{}_{I=1/2}G_{1/2}
\left\{   P_J\left(1+\frac{t}{2\Phi}\right)
   \left[ \frac{1}{\nu+(t+2\theta)}-\frac{1}{\nu-(t+2\theta)}
   \right]
\right.\ \
\nonumber \\
&   &
\ \ \ \ \ \ \ \ \ \ \ \ \ \ \ \ \ \ \ \ \ \ \ \ \ \ \ \ \ \ \ \ \
\ \ \ +
\left.
\frac{P_J\left(1-\frac{\Sigma}{2\Phi}\right)-
       P_J\left(1+\frac{t}{2\Phi}\right) }{\Sigma+t}
\right\},
\label{7.1} \\
  B(\nu,t)  = & - & 2\sum^{}_{I=1/2}G_{1/2}
          P_J\left(1+\frac{t}{2\Phi}\right)
 \left[ \frac{1}{\nu+(t+2\theta)}+\frac{1}{\nu-(t+2\theta)}
 \right]\ \ \ \
\label{7.2}
\end{eqnarray}
converge uniformly near the point $\nu_t=0,\ t=0.$
This feature allows one to rewrite them in the form of
convergent power series
\begin{equation}   \label{7.4}
A(\nu,t)=\sum_{i,j=0}^{\infty}a_{ij}\nu^it^j\ , \ \ \ \ \ \
\ \ \ \ \ \
B(\nu,t)=\sum_{i,j=0}^{\infty}b_{ij}\nu^it^j\ ,
\end{equation}
with the low energy coefficients $a_{ij}$ and $b_{ij}$
completely determined by the parameters $G_I,\ M_I,\ \sigma$
appearing in the right hand sides of (\ref{7.1}) and
(\ref{7.2}). Clearly, owing to the symmetry properties
\begin{equation}   \label{7.5}
a_{2k+1,j}=b_{2k,j}=0, \ \ \ \ \ \ \ (j,k=0,1,\ldots).
\end{equation}

Let us first consider the coefficient $b_{10}$. From
(\ref{7.2}) one obtains
\begin{equation}    \label{7.6}
b_{10}=\sum_{(I=1/2)}^{}\frac{G_{1/2}}{\theta^2}.
\end{equation}
Chiral $SU_2 \times SU_2$ symmetry tells us that at $\mu=0$ the
left hand side of (\ref{7.6}) is equal to $1/(4f^2_{\pi})$.
Thus we obtain
\begin{equation}   \label{7.7}
\frac{1}{4f^2_{\pi}}=
\sum_{(I=1/2)}^{}\frac{G_{1/2}}{M^2-m^2},
\end{equation}
where both sides should be computed at $\mu^2=0$. However, the
pion mass is very small and to get an estimate one can take the
physical values of the parameters in the rhs of the relation
(\ref{7.7}). This gives (in units of $GeV^{-2}$):
\begin{equation}    \label{7.8}
33.0=(20 \pm 2.0)+ \ldots\ .
\end{equation}
The difference between two sides in (\ref{7.8}) is too large to
be explained by the corrections connected with the pion mass.
The second line of the Table 1 shows that it also cannot be
attributed to slow convergence of SR (\ref{7.7}).

Natural solution to this problem is provided by suggestion on
the existence of a relatively light resonance (or, perhaps, two
ones) with $I=1/2$. It must be a scalar, because otherwise the
correct balance in SR (\ref{6.5}) would be disturbed. This
resonance is known as $\kappa$-meson. It appears in various
theoretical schemes as well as in results of the analysis of
experimental data (K-matrix, Pad\'{e}-approximants, etc; see
\cite{13}, \cite{12} and previous review issues by PDG). The
current status of $\kappa$-meson is even less clear than that
of $\sigma$-meson.  It should be noted, however, that --- after
a long hiatus --- interest in both particles has quickened in the
past few years.  Many authors (see \cite{14} --- \cite{14g}) have
reanalized the problem of light scalars from rather different
viewpoints (potential models, unitarized resonance models,
K-matrix analysis, chiral symmetry, etc) and concluded that light
broad scalar mesons do exist, though their parameters (masses and
coupling constants) still cannot be fixed with sufficient
accuracy. Further theoretical and experimental efforts are needed
to clarify the situation in scalar sector.

Since the SR (\ref{7.7}) follows from rather general postulates,
we can use it to estimate the $\kappa$-meson parameters. Based
on the assumption that there is only one light scalar with the
mass $M\leq 1.4~GeV$ one obtains from (\ref{7.7}) and
(\ref{7.8}) the following (rough) estimate
\begin{equation}   \label{7.9}
\frac{G_{\kappa}}{(M^2_{\kappa}-m^2)^2}
 \sim 10~GeV^{-2}\ .
\end{equation}
Using this estimate and the expression (\ref{A.19}) for
$G_{1/2}^{(0)}$ one concludes that $\kappa$-meson with
$M_{\kappa}=1~GeV$ would have the width
$ \Gamma_{\pi K} \approx 220~MeV $
while $M_{\kappa}=1.4~GeV$ would correspond to
$ \Gamma_{\pi K} \approx 1~GeV $. In what follows we assume that
\begin{equation}   \label{7.10}
M_{\kappa} \approx 1~GeV\ .
\end{equation}
This value should not be taken too seriously: it provides
only indicative numbers. Light scalar mesons -- if exist --
are broad; in such a case the very meaning of the term "width"
loses its definiteness. We imply the meaning suggested by the
relations (\ref{A.17}) --- (\ref{A.19}).

The relations (\ref{7.9}) --- (\ref{7.10}) allow one to
estimate the magnitudes of the $\kappa$-meson contributions
to numerical values of the coefficients $a_{ij}$ and $b_{ij}$.
Let us consider first the latter ones. From
(\ref{7.2}) one derives
\begin{eqnarray}
 & & b_{11}=-\sum_{(I=1/2)}^{}\frac{G^{(J)}_{1/2}}{\theta^3}
       \left[ 1- \pi^{(1)}_J \xi \right]\ ,
\nonumber \\
 & & b_{12}= \sum_{(I=1/2)}^{}\frac{G^{(J)}_{1/2}}{\theta^4}
       \left[\frac{3}{4}-\pi^{(1)}_J \xi + \pi^{(2)}_J \xi^2
       \right]\ ,
\label{7.12}      \\
 & & b_{30}=\frac{1}{4}\sum_{(I=1/2)}^{}
                  \frac{G^{(J)}_{1/2}}{\theta^4}\ ,
\nonumber
\end{eqnarray}
where
\begin{equation}
\pi_J^{(k)}\equiv \frac{1}{2^k(k!)^2}\frac{(J+k)!}{(J-k)!}\ ,
\ \ \ \ \
(k\leq J);\ \ \ \ \ \ \ \ \ \
\pi_J^{(k)}=0, \ \ \ \ \ (k > J);
\label{7.15}
\end{equation}
and
$$
\xi = \frac{\theta}{2\Phi}.
$$
The corresponding numerical values can be obtained with the
help of data \cite{12} and the estimates (\ref{7.9}),
(\ref{7.10}). They are the following
\begin{equation}
b_{11}=(53.5 \pm 10)~GeV^{-4}; \ \ \ \ \
b_{12}=(-97 \pm 11)~GeV^{-6}; \ \ \ \ \
b_{30}=(13.5 \pm 2)~GeV^{-6}.
\label{7.16}
\end{equation}
Numerical values of the individual contributions to SR
(\ref{7.12}) are shown in the Table 1. It is clear that in all
three cases the most significant contribution follows from the
lightest vector resonance --- $K^*(892)$, the influence of
$\kappa$-meson appearing mainly in the values of error bars.
Heavy mesons (those with $M>2~GeV$) play negligible role because
the sum rules under consideration possess extremely rapid
convergence. The above statements certainly remain true with
respect to higher order coefficients $b_{ij}$. Moreover, the
values of $b_{ij}$ at $j \neq 0$ --- in contrast to those of
$b_{j0}$ --- only weakly depend on the assumption (\ref{7.10}).

Thus we conclude that chiral VMD (vector meson dominance)
hypothesis works satisfactory (with accuracy $\sim 25\%$) for all
the coefficients except $b_{10}$, in which case the scalar meson
contribution represents more than $30\%$ of the total value.
However, it should be remembered that the validity of this
statement strongly depends on the suggestion (\ref{7.10}); our
conclusion would be quite different if we take
$M_{\kappa}=800~MeV$.

Let us consider now the coefficients $a_{ij}$. From the
structure of (\ref{7.1}) it follows that at $i \neq 0$ the
value of $a_{ij}$ is completely determined by the contributions
of mesons with $I=1/2$. In particular,
\begin{equation}    \label{7.19}
a_{20} =  \frac{1}{2}\sum_{I=1/2}^{}
            \frac{G^{(J)}_{1/2}}{\theta^3}\ ,
\ \ \ \ \ \ \ \ \ \ \
a_{21} =  \frac{1}{4}\sum_{I=1/2}^{}
             \frac{G^{(J)}_{1/2}}{\theta^4}
  \left[-3+\pi_J^{(1)} \frac{\theta}{\Phi}\right].
\end{equation}
The corresponding numerical values
\begin{equation}    \label{7.21}
a_{20}=(17.2 \pm 2.5)~GeV^{-4},\ \ \ \ \ \ \ \ \ \ \ \ \
a_{21}=(32.0 \pm 6.8)~GeV^{-6},
\end{equation}
are caused mainly by the contribution of $K^*(892)$ (see
Table 1), the latter conclusion being strongly connected
with the assumption (\ref{7.10}).

A consideration of the Table 1 allows one to understand the
reason for applicability of VMD hypothesis in the cases
considered above. Because of extremely rapid convergence of
SR (\ref{7.12}) and (\ref{7.19}), the most significant
contribution is provided by the lightest resonance. Since the
assumed value of $M_{\kappa}$ is larger than $M_{K^*}$, the
influence of the $\kappa-$meson happens to be weaker than
that of $K^*(892)$.

The matters are much more complicated with respect to the
coefficients $a_{0j}$. Let us compute the lowest one. From
(\ref{7.1}) we have
\begin{equation}
a_{00}=\sum_{I=0}^{}\frac{G_0}{M^2}
       P_J\left(\frac{\Sigma}{4F}\right)
     - \sum_{I=1}^{}\frac{G_1}{M^2}
       P_J\left(\frac{\Sigma}{4F}\right)
    + 2\sum_{I=1/2}^{}\frac{G_{1/2}}{\Sigma}
 \left[ P_J\left( 1-\frac{\Sigma}{2\Phi} \right)
       -\frac{\sigma}{\theta} \right] .
\label{7.22}
\end{equation}
Chiral $SU_2\times SU_2$ symmetry tells us that at $\mu =0$
\begin{equation}   \label{7.23}
a_{00}=0.
\end{equation}
The latter condition allows one to get an idea on the magnitude
of total scalar-isoscalar meson contribution. Using \cite{12}
we compute the contributions of resonances with $I=1/2$ ---
the numbers are shown in Table 1. The influence of $\kappa-$meson
is estimated just as above; it happens to be relatively less
important than in SR (\ref{7.7}). Next we compute the
contributions of isovectors $\rho (770)$ $(\sim 3.0)$ and
$\rho_3(1690)$ $(\sim 2.6)$ and spin-2 isoscalar
$f_2(1270)$ $(\sim 7.3)$. Summing all the numbers (with the most
pessimistic values of error bars) one obtains from (\ref{7.22})
and (\ref{7.23}) the following sum rule
\begin{equation}   \label{7.24}
\sum_{I=J=0}^{}\frac{G_0^{(0)}}{M^2} \approx 53.5 \pm 7.5\ .
\end{equation}
This relation clearly demonstrates that chiral VMD does not apply
to the coefficient $a_{00}$: in this case the contribution due to
scalar mesons happens to be larger than that of vector ones (cf
the numbers in Table 1 with the rhs of (\ref{7.24})). An idea of
the required structure of scalar sector can be gained from the
estimate of $f_0(1300)$ contribution. Taking (see \cite{12})
$M_f=1.25~GeV,\  \Gamma_{\pi \pi}=0.37~GeV$ and
$\Gamma_{K\bar{K}}=0.03~GeV,$ one obtains the number
$$
\frac{G_0^{(0)}(f_0)}{M^2_{f_0}}\approx 3.2\ ,
$$
which is negligibly small compared to that required by
(\ref{7.24}). This estimate shows that SR (\ref{7.24}) requires
the existence of light scalar-isoscalar resonance strongly
coupled to both $\pi \pi$- and $K\bar{K}$-channels. In principle,
the mentioned above $\sigma$-meson would be a good candidate for
this role. If we take this hypothesis (along with the parameters,
taken from the quoted above papers \cite{14} --- \cite{14g}), the
computation of the coefficients $a_{0j}$ could be easily done.
However, it should be kept in mind that the same resonance
appears also in processes of $\pi \pi$ and $K\bar{K}$
scattering. Therefore, from the purely theoretical point of
view, it is much more interesting to carry out simultaneous
analysis of joint system of sum rules in order to get
self-consistent results. This analysis is in progress now.

\section{Concluding remarks}
\mbox{}

The method of Cauchy's forms described in Sec.~3 allowed us to
avoid model dependence of results. Our conclusion concerning the
dual properties of tree-level amplitude in effective field theory
with infinite number of field species follows directly from
certain analyticity requirements (meromorphy and polynomial
boundedness) and the requirement of crossing-symmetry. The same
is true with respect to the system of bootstrap constraints
which appears as necessary and sufficient condition providing
feasibility of analytic continuation between the direct- and
cross-channel domains. This conclusion eliminates apparent
contradiction between the conventional quantum field theory
(QFT) approach and that based on the idea of duality (in Ref.
\cite{16}
this problem is considered from different point of view). Dual
amplitude is constructed from the infinite series of direct
channel poles, the cross channel ones appearing just as a result
of summation of this series and its subsequent analytic
continuation to the cross-channel domain. In contrast, the Born
approximation in QFT contains both types of poles simultaneously
(plus smooth -- background -- terms corresponding to point-like
interactions). This very feature is commonly considered as
drastic difference between two approaches.

Our results show that there is logical gap in the above
reasoning. Indeed, the dual amplitude contains infinite
number of poles corresponding to the states with arbitrarily
high values of spin and mass. Therefore, it is natural to compare
it with QFT which also contains the infinite spectrum of bound
states. Next, as it follows from our analysis in Sec.~4, every
dual amplitude along with infinite set of poles contains also
specific background terms which manifest themselves explicitly
in the corresponding area of the momentum transfer. Thus the
QFT in question can also contain pointlike interaction terms.
At last, the QFT amplitude -- as well as the dual one --
should be written in the correct analytic form, {\it the latter
one depending on the domain under consideration}. In particular,
in the direct-channel domain this form cannot contain any poles
in momentum transfer, the contribution of
$t$-channel exchange graphs looking here like the background
interaction with infinite number of derivatives. With this
understanding in mind, one can write down the most general QFT
expression for tree-level amplitude (which is nothing but the
amplitude of the effective QFT) in the form dictated by the
analyticity requirements, and then carry out the analytic
continuation to the cross-channel domain. The conditions
guaranteeing that the resulting expression will contain no other
singularities but simple poles and ambiguity points, and
possess the desired asymptotic behavior, are precisely
those expressed by the system of bootstrap constraints.

Thus we conclude that the effective field theory of strong
interactions, based on the idea of quark-hadron duality,
necessarily results in the string form of tree-level amplitude
provided that certain analyticity requirements (meromorphy and
polynomial boundedness) are imposed. This conclusion provides a
solution to the problem of string organization of field theories
\cite{16}.

Further, though the mere appearance of bootstrap constraints does
not depend of the values of bounding polynomial degrees, their
particular form does depend of those values. It is remarkable
that numerical test based solely on low energy data provides
clear arguments in favor of the degrees corresponding to
experimentally known values of intercepts. This means that the
formulae for low energy coefficients (Sec.~7), obtained as
by-product of our study of effective hadron theory with maximal
analyticity, may be thought of as model independent results based
on well established general principles. However, it should be
kept in mind that, to use those formulae in ChPT computations,
one needs to expand them in powers of quark masses. This is
necessary just to avoid contradiction with chiral power counting
rules.

Our main conclusion concerning the structure of LEC's is the
following. The idea of Chiral duality
\cite{4,5,18,19}
(for the review see
\cite{20,21})
is certainly true. It mirrors the requirements of general
principles of quark-hadron duality and analyticity. Thus it is
no less general than ChPT itself. However, this idea needs more
accurate formulation. Indeed, a comparison of the well-defined
form
(\ref{7.1})
with the formally written expression
(\ref{A.13})
shows considerable difference in their structure. In contrast
with
(\ref{A.13}),
the expression
(\ref{7.1})
does not contain any unknown polynomials like
$E_A(s,t,u)$.
Instead, it contains the contribution (see the second item)
depending on the parameters of isovector resonances; such a term
could not appear in the "naive" form
(\ref{A.13})
in principle. On the other hand, the well-defined form
(\ref{7.2})
does not contain any contribution from isovectors, while the
formal expression
(\ref{A.14})
does contain it (along with unknown polynomial
$E_B(s,t,u)$
which is absent in
(\ref{7.2})).
This means that one should exercise caution when formulating the
idea of Chiral duality. In this respect, the situation resembles
that with formulation of VMD hypothesis
\cite{18,19,22} ---
the latter happened to be well defined only under the condition
if certain limitations are imposed on high energy asymptotics of
the vector meson contribution. In fact, our polynomial
boundedness requirement is nothing but a generalized version of
those limitations applied to the full tree-level amplitude.

In contrast to Chiral duality, the status of Chiral VMD
hypothesis is much less reliable. From recent data analysis and
from our SR
(\ref{7.22})
it follows that the existence of
{\it light}
broad scalar resonance looks necessary to explain the low energy
experimental data. In some cases (like
$a_{0j}$ --
see Sec.~7) the contribution of this scalar meson may happen to
be significant even compared to (also allowed) that of lightest
vector mesons. Modern understanding of the scalar sector still
looks unsatisfactory.

Here it is a point to stress the difference between our approach
and that used by those authors who study various QCD-inspired
models (for the review, see
\cite{23,24})
to compute chiral LEC's. Our results show that LEC's can be
treated as "secondary" quantities completely fixed by the values
of "primary" ones: hadron masses and on-shell coupling
constants. The latter values should be taken from the underlying
fundamental theory (QCD, string, ...). Given them one can
compute all other characteristics of low-energy hadron reactions
in a framework of the approach based on effective field theory
accounting for few general principles (symmetry and analyticity).
In contrast, the authors of QCD-inspired models consider LEC's on
the same ground as spectrum parameters. Particular assumptions
(inavoidable in this approach) concerning the hadronization
regime in QCD introduce strong model dependence in the results.
This feature along with scarcity of modern database (specially
stressed with respect to
$(\pi, K)$
processes in
\cite{25})
hampers the understanding of relative importance of different
mechanisms. That is why we prefer to use the conventional
approach
\cite{3,4,5},
supplemented with requirements imposed by the general principle
of maximal analyticity.

It should be noted that the latter principle plays no role (or,
better, it is trivial) in conventional renormalizable field
theories: there is no necessity to postulate anything which can
be computed. The necessity of considering maximal analyticity as
independent condition appears only if the number of field species
(or, the number of derivatives) is allowed to become infinite.
As shown recently in
\cite{27},
in this very case one can expect considerable simplification of a
theory near the phase transition point. Therefore, it would be
interesting to find an algebraic structure corresponding to the
considered above infinite system of bootstrap constraints. It is
more or less clear that it might be one of the algebras of
rational functions. This suggestion correlates (though
indirectly) with the structure of our sum rules which admit
existence of infinite-dimensional multiplets.

\section*{Acknowledgments}
\mbox{}

It is a pleasure to thank A.Andrianov, H.V. von Geramb and
M.D.Scadron for the information on recent publications and
J.Gasser  for his interest to this work and friendly
support. We are indebted to M.Jarmolovich for her help in
preparation of the manuscript.

This work was supported in part by RFBR (Grant 98-02-18137) and
by GRACENAS (Grant 6-19-97, 1997). The work of A.Vereshagin was
supported also by ISSEP "Soros Students" (Grant s97-2391).

\section*{Appendix}
\mbox{}

Here we give a summary of formulae and relations which are
necessary for the analysis of $(\pi, K)$ processes.

Three different channels of the reaction under consideration
are the following
$$
\begin{array}{ccccccc}
\pi_a(k_1) & + & K_{\alpha}(p_1) & \longrightarrow &
\pi_b(k_2) & + & K_\beta(p_2)\ ,  \\
\pi_a(k_1) & + & \pi_b(k_2)  & \longrightarrow &
\overline{K_\alpha}(p_1) & + & K_\beta(p_2)\ ,  \\
\pi_a(k_1) & + & \overline{K_{\beta}}(p_2) & \longrightarrow &
\pi_b(k_2) & + & \overline{K_\alpha}(p_1)\ .
\end{array}
$$
Here $a,b=1,2,3$ and $\alpha,\beta=1,2$ stand for isotopic
indices. The amplitude can be written as follows
\begin{equation}     \label{A.4}
M_{\beta\alpha}^{ba} = \delta_a^b\delta_\alpha^\beta A(s,t,u)
  + i\varepsilon_{bac}(\sigma_c)_{\beta\alpha} B(s,t,u)\ ,
\end{equation}
where $ Tr(\sigma_a\sigma_b)=2\delta_{ab}\ $, and
$$
\  s=(k_1+p_1)^2,\ \ \ \ t=(k_1-k_2)^2,\ \ \ \ u=(k_1-p_2)^2,
$$
$$
s+t+u=2(m^2+\mu^2)\equiv2\sigma.
$$
Here $\mu\ (m)$ is the pion (kaon) mass.
Due to requirements of Bose symmetry
\begin{equation}
A(s,t,u)=A(u,t,s); \ \ \ \ \ \ \ \
B(s,t,u)=-B(u,t,s)
\label{A.7}
\end{equation}
We use also 3 different pairs of independent kinematical
variables $\{\nu_x,x\}\ \  (x=s,t,u): $
$$
\nu_s=u-t;\ \ \ \ \ \nu_t=s-u;\ \ \ \ \ \nu_u=t-s.
$$
Each set $\{\nu_x,x\}$ forms a natural coordinate system in the
3-dimensional band $B_x$ corresponding to small real $x$ and
arbitrary complex $\nu_x$, the section of $B_x$ by Mandelstam
plane (real s,t,u) resulting in a 2-dimensional strip $S_x$
parallel to the side x=0 of the Mandelstam triangle (see Fig.1).

The Regge theory prescriptions for the asymptotic behavior of
the amplitudes $A$ and $B$ in the bands $B_s, B_t, B_u$
can be summarized as follows:
\begin{eqnarray}
B_s \{|\nu_s|\rightarrow \infty ;\ s \sim 0 \}:\ \ \
&
\left\{
\begin{array}{rcll}
\left.(A+2B)\right/_{B_s}\
& \sim
& \nu_s^{\alpha_{1/2}(s)};\ \ \ \ \ \ \
& [N=0];        \\
\left.(A-B)\right/_{B_s}\ \
& \sim
& \nu_s^{\alpha_{3/2}(s)};\
& [N=-1];
\end{array}
\right.
\label{A.9}     \\
B_t \{|\nu_t|\rightarrow \infty ;\ t \sim 0 \}:\ \ \
&
\left\{
\begin{array}{rcll}
\ \ \ \ \ \ \ \ \ \
\left.A\right/_{B_t}\
& \sim
& \nu_t^{\alpha_0(t)};\ \ \ \ \ \ \ \ \
& [N=0];        \\
\left.B\right/_{B_t}\
& \sim
& \nu_t^{\alpha_1(t)};\
& [N=-1];
\end{array}
\right.
\label{A.10}    \\
B_u \{|\nu_u|\rightarrow \infty ;\ u \sim 0 \}:\ \ \
&
\left\{
\begin{array}{rcll}
\left.(A-2B)\right/_{B_u}\
& \sim
& \nu_u^{\alpha_{1/2}(u)};\ \ \ \ \ \ \
& [N=0];        \\
\left.(A+B)\right/_{B_u}\ \
& \sim
& \nu_u^{\alpha_{3/2}(u)};
& [N=-1].
\end{array}
\right.
\label{A.11}
\end{eqnarray}
Here we also show in braces the degrees of bounding polynomials
needed to construct the corresponding Cauchy forms. Those
degrees are uniquely determined by the known intercepts of the
leading Regge trajectories with the isospin $I$:
$$
\alpha_0(0)=1;\ \ \ \ \alpha_1(0)\approx0,5;\ \ \ \
\alpha_{1/2}(0)\approx0,3;\ \ \ \ \alpha_{3/2}(0)<0.\ \
$$

For the sake of the reader's convenience, below we give also the
\underline{formal} (i.e. constructed in accordance with "naive"
Feynman rules) tree-level expressions for the effective
amplitudes $A$ and $B$ appearing in (\ref{A.4}).
\begin{eqnarray}
A(s,t,u)&=&-\sum_{I=0}^{}G_0\frac{P_J(\frac{s-u}{4F})}{t-M^2} -
\sum_{I=1/2}^{}G_{1/2}P_J(1+\frac{t}{2\Phi})
\left\{ \frac{1}{s-m^2} + \frac{1}{u-m^2}\right\}\ \ \ \ \ \
\nonumber   \\
        && + E_a(s,t,u).
\label{A.13}   \\
B(s,t,u)&=&-\sum_{I=0}^{}G_1\frac{P_J(\frac{s-u}{4F})}{t-M^2} -
\sum_{I=1/2}^{}G_{1/2}P_J(1+\frac{t}{2\Phi})
\left\{ \frac{1}{s-m^2} - \frac{1}{u-m^2}\right\}
\nonumber   \\
        && + E_B(s,t,u).
\label{A.14}
\end{eqnarray}
Here $E_A$ and $E_B$ stand for the \underline{formal} power
series in $s,t,u$ obeying the Bose symmetry conditions
(\ref{A.7}), and
$$
F\equiv F(M^2,m^2,\mu^2)=
\frac{1}{4}\left|\sqrt{(M^2-4m^2)(M^2-4\mu^2)}\right|\ ,
$$
$$
\Phi\equiv \Phi (M^2,m^2,\mu^2)=
\frac{1}{4M^2}
\left|\sqrt{M^4+m^4+\mu^4-2M^2m^2-2M^2\mu^2-2m^2\mu^2}\right|\ .
$$
The explicit formulae expressing the constants $G_I$
in terms of the corresponding decay widths look as follows
\begin{equation}
\left|G_0^{(J)}\right|=8\pi M^2_S (2J+1)
\sqrt{\frac{1}{3}
\frac{\Gamma(S\rightarrow\pi \pi)}{|\vec{p}_{\pi}|}
\frac{\Gamma(S\rightarrow K\bar{K})}{|\vec{p}_{\tt K}|}}\ , \ \ \ \
(M_S\geq 2m);
\label{A.17}
\end{equation}
\begin{equation}
\left|G_1^{(J)}\right|=8\pi M^2_V (2J+1)
\sqrt{\frac{1}{2}
\frac{\Gamma(V\rightarrow\pi \pi)}{|\vec{p}_{\pi}|}
\frac{\Gamma(V\rightarrow K\bar{K})}{|\vec{p}_{\tt K}|}}\ , \ \
\ \  (M_S\geq 2m);
\label{A.18}
\end{equation}
\begin{equation}   \label{A.19}
G_{1/2}^{(J)}=8\pi M^2_R (2J+1)\frac{1}{3}
\frac{\Gamma (R\rightarrow \pi K)}{|\vec{p}|}\ , \ \ \ \ \ \
\ \ \ \ \ \
(M_R\geq m+\mu).
\end{equation}
As in the Ref. \cite{1} we use the notations
$$
D_s=B_t\cap B_u\ ;\ \ \ \ \ D_t=B_u\cap B_s\ ;\ \ \ \ \
D_u=B_s\cap B_t
$$
for the mutual domains of various pairs of the bands $B_x$ and
$$
\theta \equiv M^2 - \sigma;\ \ \ \ \ \ \ \
\Sigma\equiv M^2 - 2\sigma
$$
for two special combinations of masses (here
$M$ stands for the resonance mass).

\begin{table}

\begin{center}
\begin{tabular}{||c||l|c|l|c|l|l|l|l||}  \hline \hline
meson & ~~\( \kappa \) & \( K_0^* \) &
~~\( K^* \) & \(K^* \) & ~~\(K^* \) &
\( K_2^* \) & ~~\(K_3^* \) & \( K_4^* \) \\
mass & ~1.0 & ~1.43 & ~0.89 & ~1.41 &
~1.68 & 1.43 & ~1.78 & 2.05 \\  \hline \hline

rhs(\ref{6.2}) & ~~~-- & --& ~24.0   & ~0.54 &
~1.31 & 2.60  & ~0.60 & 0.28 \\  \hline

\( b_{10} \)   & ~~10 & ~2.36 & ~12.1 & ~0.42 &
~1.75 & 2.18 &~0.87 & 0.43 \\ \hline

\( b_{11} \)   & ~--13  & --1.30 & ~48.9  & ~0.31 &
~0.76 & 7.17 & ~3.67 & 2.23 \\ \hline

\( b_{12}\)    & ~~12 & ~0.55 & --103 & --0.22 &
--0.36 & 1.25 & ~2.50 & 2.30 \\ \hline

\( b_{30}\)    & ~~3.6 &  ~0.18 & ~10.8 & ~0.03 &
~0.07 & 0.17 & ~0.03 & 0.01 \\ \hline \hline

\( a_{20}\)    & ~~5.4 & ~0.65 & ~11.42 & ~0.12 &
~0.34 & 0.60 & ~0.15 & 0.05 \\ \hline

\( a_{21}\)    & ~--11 & --0.55 & ~35.2 & ~0.06
& ~0.08 & 1.83 & ~1.29 & 0.57 \\ \hline

\( a_{00}\)    & --4.12 & --1.47 & --27.7  & --1.87 &
--9.74 &   6.91 & --6.40 &   3.00 \\ \hline
\end{tabular}
\end{center}

\caption{Separate contributions of \( I = \frac{1}{2} \)
         mesons to SR (\ref{6.2}) and LEC's.}

\end{table}

\end{document}